# High-Throughput Discovery of Kagome Materials in Transition Metal Oxide Monolayers

Renhong Wang (王人宏)[1,2], Cong Wang (王聪)[1,2,†], Ruixuan Li (李睿宣)[1,3], Deping Guo (郭的坪)[4,1], Jiaqi Dai (戴佳琦)[1,2], Canbo Zong (宗灿波)[1,2], Weihan Zhang (张伟涵)[1,2], and Wei Ji (季威)[1,2,†]

[1] *Beijing Key Laboratory of Optoelectronic Functional Materials & Micro-Nano Devices, School of Physics, Renmin University of China, Beijing 100872, China*
[2] *Key Laboratory of Quantum State Construction and Manipulation (Ministry of Education), Renmin University of China, Beijing 100872, China*
[3] *Beijing No.35 High School, Beijing 100037, China*
[4] *College of Physics and Electronic Engineering, Sichuan Normal University, Chengdu 610101, China*

**ABSTRACT:** Kagome materials are known for hosting exotic quantum states, including quantum spin liquids, charge density waves, and unconventional superconductivity. The search for kagome monolayers is driven by their ability to exhibit neat and well-defined kagome bands near the Fermi level, which are more easily realized in the absence of interlayer interactions. However, this absence also destabilizes the monolayer forms of many bulk kagome materials, posing significant challenges to their discovery. In this work, we propose a strategy to address this challenge by utilizing oxygen vacancies in transition metal oxides within a "1+3" design framework. Through high-throughput computational screening of 349 candidate materials, we identified 12 thermodynamically stable kagome monolayers with diverse electronic and magnetic properties. These materials were classified into three categories based on their lattice geometry, symmetry, band gaps, and magnetic configurations. Detailed analysis of three representative monolayers revealed kagome band features near their Fermi levels, with orbital contributions varying between oxygen 2*p* and transition metal *d* states. This study demonstrates the feasibility of the "1+3" strategy, offering a promising approach to uncovering low-dimensional kagome materials and advancing the exploration of their quantum phenomena.

† Corresponding author. E-mail: wcphys@ruc.edu.cn
† Corresponding author. E-mail: wji@ruc.edu.cn

## 1. Introduction

Kagome lattices, composed of corner-sharing triangles arranged in hexagonal symmetry, exhibit unique electronic properties such as Dirac cones, van Hove singularities, and flat bands. These electronic characteristics provide an versatile platform for exploring a wide range of quantum states[1, 2], such as a candidate for quantum spin liquid state in $ZnCu_3(OH)_6Cl_2$[3, 4], Weyl semimetal state in $Co_3Sn_2S_2$[5, 6, 7, 8, 9], unconventional superconductivity and charge density wave states in $AV_3Sb_5$ (A=alkali metals)[10, 11, 12, 13, 14, 15, 16]. However, the kagome layers in these bulk materials are usually hybridized with adjacent non-kagome layers, which results in the characteristic kagome bands being complicated with dense bulk bands[14, 7] and often residing away from the Fermi level[17]. With largely suppressed interlayer couplings and minimal interference from other bands, two-dimensional (2D) kagome materials offer promising opportunities to isolate pristine kagome bands near the Fermi level[18, 19, 20, 21, 22]. Several methods have been explored to construct 2D kagome materials. One straightforward approach is direct exfoliation from bulk kagome materials[18–21]; however, overcoming non-van-der-Waals interlayer interactions remains challenging[8, 23]. Moreover, the presence of residual capping atoms or layers prevents effective isolation of neat kagome bands near the Fermi level in, e.g., monolayer $AV_3Sb_5$[24]. Although substantial progress has been made in moiré bi- or few-layers and surface supported kagome layers, these approaches face critical issues to solve for practically constructing neat kagome bands[25, 26, 27, 28, 19, 29, 30, 31]. A recently developed strategy leverages material defects to construct pristine kagome monolayers. For instance, mirror twin boundaries (MTBs) are preferably formed and orderly aligned in uniform lattices at a certain range of Te chemical potential in single-layer $MoTe_{2-x}$[22]. This process has enabled the formation of various types of kagome structures, such as $Mo_5Te_8$[21] and $Mo_{33}Te_{56}$[32].

Oxygen vacancies are commonly observed in transition metal oxides (TMOs), providing a promising avenue for constructing kagome monolayers via a defect-based strategy. Synthesis of two-dimensional metal oxide layers has been rapidly developing that monolayer metal oxides were recently reported in SnO[33] and $PtO_x$[34] although they were prepared on substrates like $SiO_2$ and Pt. A 2×2 supercell of a MO monolayer in a triangular lattice contains four formula units. If one of these units is modified by, for instance, through the formation of a vacancy, the remaining three units form a kagome lattice, referred to as the "1+3" strategy, as illustrated in Fig. 1(a). We initiated our study with a 1T-phase TMO monolayer ($MO_2$), where a single layer of metal atoms is sandwiched between two oxygen atomic layers. From this structure, four potential kagome monolayer phases were designed, as shown in Fig. 1(b)-1(d) and S1. The removal of one oxygen atom from the 2×2 supercell ($MO_{1.75}$) results in a kagome latticed structure for that layer, termed "kagome-single" (KS), as illustrated in Fig. 1(c). Further removal of one additional oxygen atom from the other oxygen sublayer ($MO_{1.5}$) induces a kagome structure of oxygen atoms also in the other oxygen sublayer, denoted as "kagome-bilayer" (KB). The second removal has two options, namely the O atom sitting at the VU1 and VU2 sites, denoted with -KB1 and KB2, respectively (Fig. S1(e) and S1(f)). Finally, the last kagome phase was constructed by removing one metal atoms, forming a $M_3O_8$ monolayer, as illustrated in Fig. 1(d).

By following this "1+3" strategy, here, we implemented a high-throughput density functional theory calculation workflow to predict thermodynamically stable kagome lattices in TMO monolayers[35, 36]. All transition metal elements, except for lanthanides, actinides and radioactive element Tc, were considered, among which we identified 12 thermodynamically stable kagome monolayers. Their electronic structures and magnetic properties were studied and summarized, based on which these 12 monolayers were categorized into three groups. We conducted a detailed analysis on three representative monolayers, including Sc-KB1, $Ta_3O_8$, and $Ir_3O_8$. We also found that the orbital composition of kagome bands significantly depends on the residual valence electron count of the transition metals in these TMO monolayers.

## 2. **Computational method**

Our DFT calculations were carried out using the generalized gradient approximation for the exchange-correlation potential, the projector augmented wave method and a plane-wave basis set as implemented in the Vienna ab-initio simulation package (VASP)[37]. The PBE functional[38] was used to describe the exchange and correlation energy density function, and dispersion correction was implemented using the DFT-D3 method[39]. Each supercell used in our calculations contains a 20 Å vacuum layer to suppress image supercell coupling. For both geometric relaxations and electronic structure calculations, a kinetic energy cut-off of 700 eV was used, and a uniform k-point grid of density 8.0/$Å^{-1}$ was used to sample the first Brillouin zones. The energy convergence criterion of the self-consistency was set at $1 \times 10^{-5}$ eV. The shape and in-plane lattice area of each supercell were fully optimized, allowing all atoms to relax until the residual force per atom was below $1 \times 10^{-2}$ eV/Å. A uniform set of $U = 3.0$ eV and $J = 0.0$ eV[40] was used to consider on-site Coulomb interactions on transition metal atom *d* orbitals for every considered monolayer. The pseudopotentials used are as follows: for Sc, Ti, V, Y, Zr, Nb, Mo, and W, semi-core *s*- and *p*-electrons were included; for Cr, Mn, Ru, Rh, Hf, and Ta, semi-core *p*-electrons were included; and for all other transition metal elements, valence electrons were counted starting from the *d*-orbitals. A finite difference method was used to calculate phonon spectra of these monolayers at the Gamma point, with a kinetic energy cut-off of 700 eV and a uniform k-point grid of density 14.3/$Å^{-1}$. The self-consistency convergence criterion was set at $1 \times 10^{-7}$ eV and the ionic relaxation convergence criterion was set at $5 \times 10^{-4}$ eV/Å.

Now, we introduce the high-throughput workflow used in this work. There are three steps in the workflow, as shown in Fig. 2. These three steps are represented by blue, orange and green sections, respectively. In the first step, for all considered transition metal elements, both the 1T-phase and the four kagome phases were examined. As structural energy differences are typically more significant than those associated of magnetic energy differences, the ferromagnetic (FM) configuration was thus used as the initial magnetic configuration when comparing relative energies of magnetic monolayers. For each transition metal element, the formation enthalpy of each phase was calculated under oxygen rich limit and oxygen poor limit. The thermodynamic relative stability between different phases was compared to determine the most favorable phase under different oxygen chemical potentials, then we obtain the kagome structures locally stable. In the second step, for the elements with locally stable kagome structures, we additionally calculated the formation enthalpy of 11 other phases. Combined with the 5 phases considered in the first step, a total of 16 phases were examined. Repeat the process of the first step, we identified globally stable kagome TMO monolayers under different oxygen chemical potentials. In the third step, for each of the globally stable kagome TMO monolayer, the magnetic ground state was determined.

For a monolayer where its oxygen atoms form a kagome lattice, the transition metal atoms form a triangular lattice. The magnetic configurations considered for such monolayer include nonmagnetic (NM), ferromagnetic (FM), stripe antiferromagnetic (sAFM), double stripe antiferromagnetic (dAFM), and zigzag antiferromagnetic (ZZ) configurations, as shown in Fig. S3. If a kagome lattice was comprised of transition metal atoms, the inherent spin frustration complicates consideration of AFM configurations. For instead, ferrimagnetic (FiM) configurations (shown in Fig. S4) are examined for this category.

The Atomic-orbital Based Ab-initio Computation at USTC (ABACUS) package[41, 42] and PYATB[43] were used to calculate the $Z_2$ invariants and Chern number. For the topological calculations, the wavefunctions were expanded using localized atomic orbital basis sets. The atomic orbital configurations for the relevant elements are as follows: O (2s, 2p, 1d), Sc (4s, 2p, 2d, 1f), and Ta (4f, 2p, 2d, 2f, 1g). The number of atomic orbitals for each element is 5 for O, 9 for Sc, and 11 for Ta. A cutoff of 150 Ry was employed, and a self-consistency convergence criterion was set to $1\times10^{-7}$ Ry.

## 3. Results and discussion

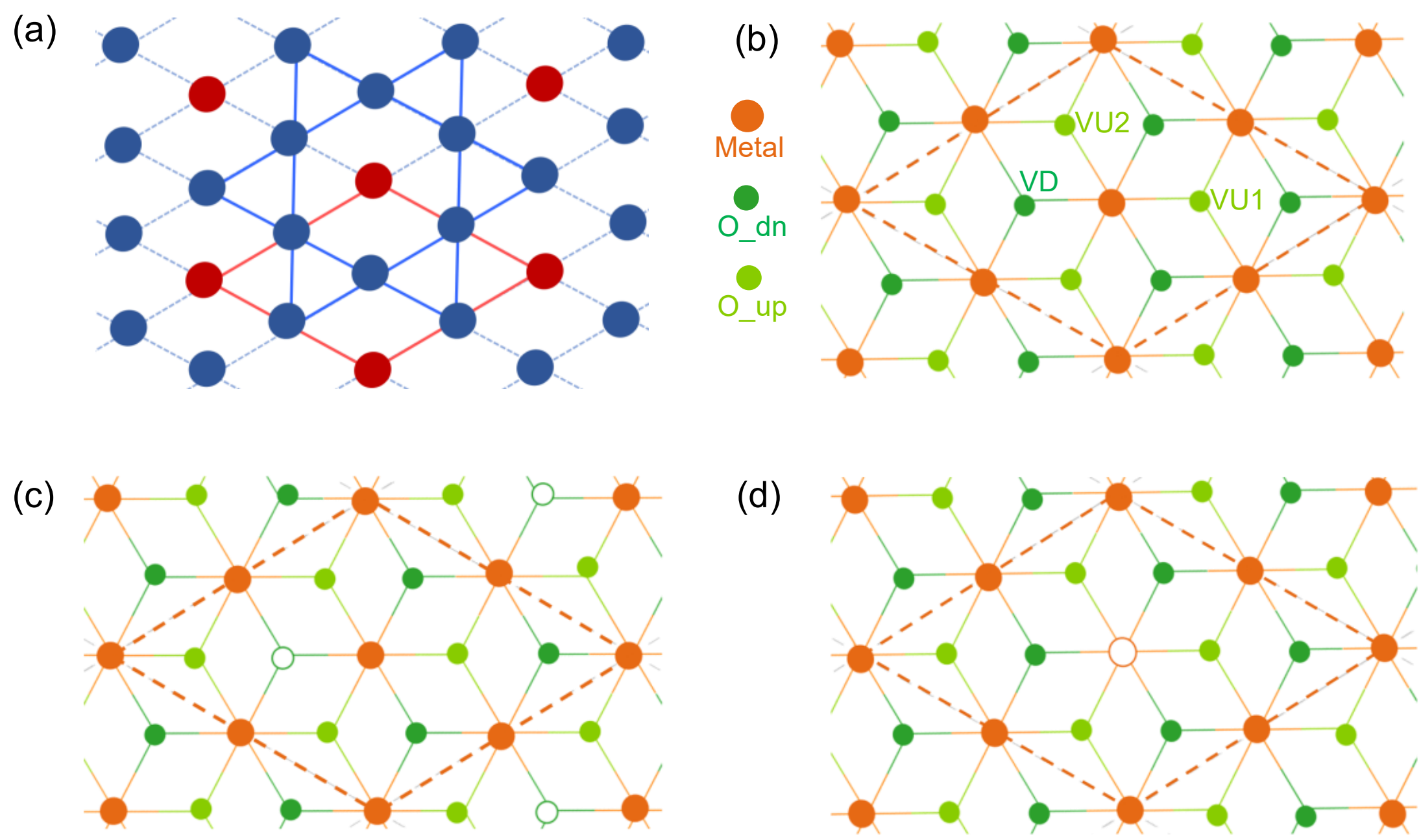

**Fig. 1.** Schematic representation of the "1+3" strategy, the 1T-phase of TMO, and potential kagome phases. (a) Illustration of the "1+3" strategy. The red lines outline a 2×2 supercell of triangular lattice, while the blue lines depict

a kagome lattice. The red and blue circles indicate distinct formula units. (b) Top view of the 1T-phase of TMO. The orang dashed lines denote the 2×2 supercell. The orange circles represent transition metal atoms, while the dark green circles and light green circles correspond to oxygen atoms in the upper and lower oxygen layers, respectively. The three possible oxygen vacancy sites are labeled VD, VU1, and VU2. (c) Top view of the "kagome-single" (KS) phase. (d) Top view of the $M_3O_8$ phase.

After the second step of the workflow (Fig. 2 orange), we filtered out 19 kagome structures that could be locally stable for each metal element. Their energies were then compared with those of 11 other phases, including four hexagonal, two triangular, and five tetragonal lattices (Fig. 2 green). These totally 16 phases cover all known potential forms ever found for transition metal oxide layers (Fig. S2). After the comparison, we obtained 12 globally stable kagome monolayers, which were denoted using a M-STRU format where M represents the transition metal and STRU denotes the structure of phase. For instance, Sc-KB1 represents $ScO_{1.5}$, where two oxygen vacancies form at the VD and VU1 sites. The upper right corner of Fig. 2 lists which elements under what oxygen chemical potential are stable in each kagome phase. Among them, Sc-KB1, Cu-KB2 and $Ta_3O_8$ are globally stable span the whole range of the O chemical potential. In the rest, seven of them are the most stable phase under the O-rich condition and V-KS and Y-KB1 have superior stability under O-deficient conditions. We also calculated the phonon spectra at the Gamma point for the 12 kagome structures and found no imaginary frequencies, indicating their potential kinetic stability (Table S2).

To explore the influence of different Hubbard U values on thermodynamic stability, we reevaluated the thermodynamic stability of the 12 globally stable kagome monolayers with $U_{\mathrm{eff}}$=2 and 4 eV. In our calculations, the predicted most, second-most and third-most stable phases were considered. As summarized in Table 1, except $Ir_3O_8$ and Cu-KB2 with $U_{\mathrm{eff}}$=4 eV, all other kagome monolayers are "robust" with $U_{\mathrm{eff}}$=2 and 4 eV. Here, the "robustness" means two folds: 1) the most stable phase changes, but, to another kagome phase (for example, the Sc-KS and Sc-KB1 monolayers); 2) the originally predicted monolayer is still the most stable one (for example, $Ta_3O_8$). Overall, all kagome phases are "robust" with a decreased $U$ value and only two predicted monolayers fail to maintain kagome phases when the $U$ value increases to 4 eV. For the 5 monolayers that exhibit kagome features near the Fermi level, the variation in U values does not affect the kagome characteristics in the band structures, as illustrated in Fig S6~S10.

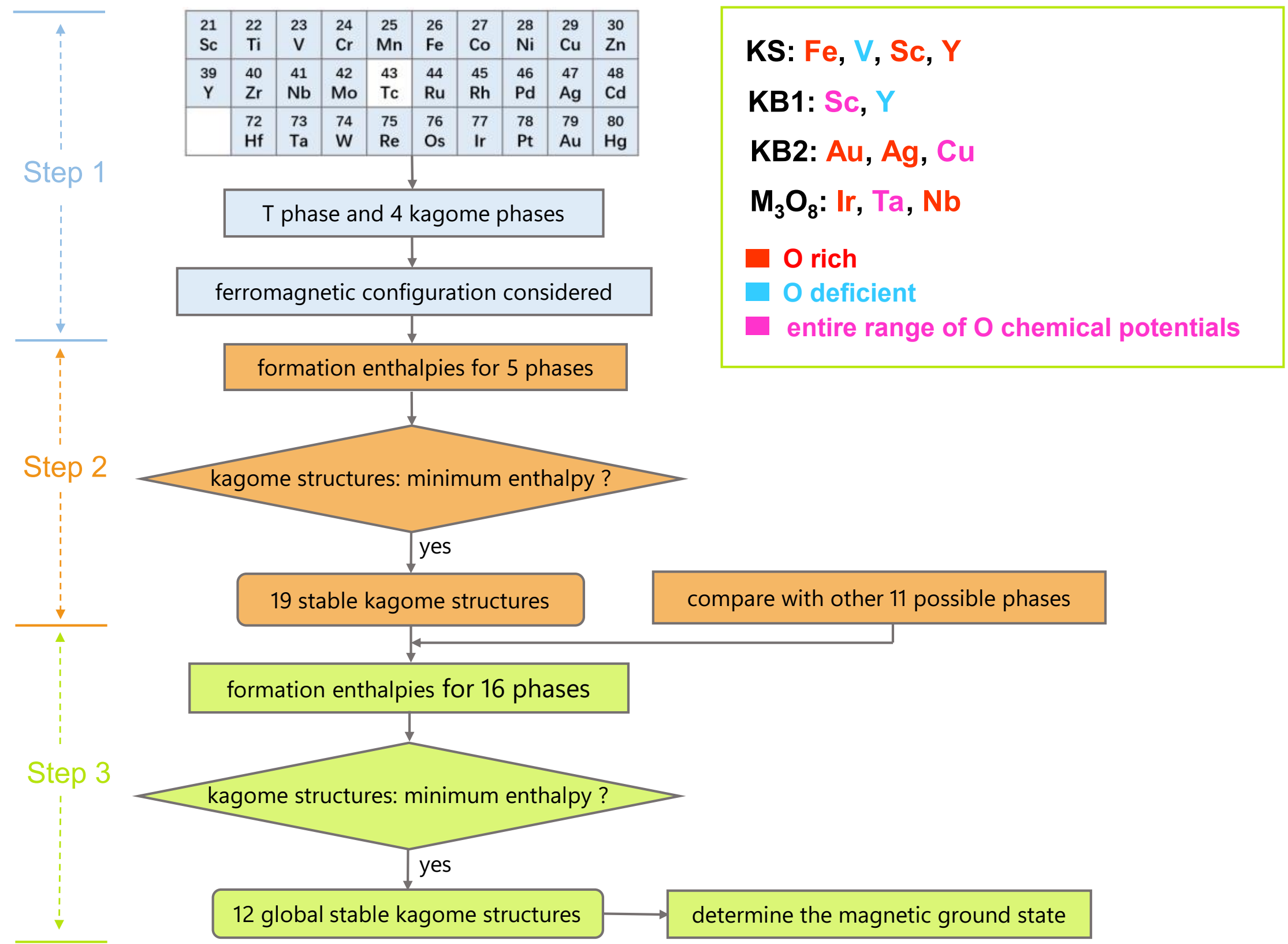

**Fig. 2.** Workflow for high-throughput prediction of kagome materials in TMO monolayers. The workflow consists of three steps, represented by blue (step 1), orange (step 2), and green (step 3) sections. The 12 globally stable kagome monolayers were listed in the upper-right green box. The black texts indicate the phases, while the red, blue, and pink texts indicate the elements that are stable under oxygen-rich, oxygen-poor, and all ranges of oxygen chemical potential, respectively.

Table 1. List of Hubbard U dependence of the thermodynamic stability for the most stable and two metastable phases under two oxygen chemical potential (CP) limits. A checkmark ( √ ) indicates that the most stable phase remains the same as under the U=3 eV condition. A circle (○) denotes a transition to another kagome phase. For cases where the monolayer does not retain the kagome phase, the most stable phase is marked.

| CP | ML | U=2 (eV) | U=4 (eV) |
| --- | --- | --- | --- |
| O-rich | $Ir_3O_8$ | √ | $IrO_2$-T |
| | Fe-KS | ○ | ○ |
| | Sc-KS | ○ | √ |
| | Sc-KB1 | √ | ○ |
| | Y-KS | √ | √ |
| | Au-KB2 | √ | √ |
| | Ag-KB2 | √ | √ |
| | Cu-KB2 | √ | √ |
| | $Ta_3O_8$ | √ | √ |
| | $Nb_3O_8$ | √ | √ |
| | Cu-KB2 | √ | CuO-F |
| | V-KS | ○ | ○ |

| O-deficient | Sc-KB1 | √ | √ |
|---|---|---|---|
| | Y-KB1 | √ | √ |
| | $Ta_3O_8$ | √ | √ |

The last step for this high-throughput workflow lies in the determination of their magnetic ground states. Among the 12 globally stable structures, two are non-magnetic monolayers without spin polarization (See Table 2). Three of them, including M-KB2 (M = Cu, Ag, Au), have local magnetic moments on both metal (0.53-0.80 $\mu_B$/atom) and O (0.16-0.21 $\mu_B$/atom) atoms. However, weak charge density overlaps observed between the transition metal atoms and their neighboring O atoms (Fig. S11(a)-11(c)) result in negligible magnetic exchange interactions, not preferring any magnetic order. There are four M-KS monolayers exhibiting magnetism. The local magnetic moments primarily reside around the metal atoms for Fe-KS (4.0 $\mu_B$/atom) and V-KS (1.9 $\mu_B$/atom), the magnetic orders of which are ZZ (Fig. S11(d)) and dAFM (Fig. S11(e)), respectively. In the Sc-KS and Y-KS monolayers, the magnetic moments are dominantly contributed by O atoms (0.31 and 0.34 $\mu_B$/atom), showing FM orders. All three $M_3O_8$ monolayers prefer the FM order, but the local moments are found on O atoms solely in $Nb_3O_8$ and $Ta_3O_8$ (0.16 $\mu_B$/atom for both), while the moments are contributed by both Ir (0.47 $\mu_B$/atom) and O (0.12 $\mu_B$/atom) atoms in $Ir_3O_8$.

To verify our original idea that the magnetic energy difference plays a minor role in determining the thermodynamic stability of these monolayers, in Table S3, we list the formation enthalpy differences between the most and second-most stable monolayers, and the energy differences between the FM and the magnetic ground state (GS) for each transition metal. Except for Fe-KS, the formation enthalpy difference for each transition metal element is one or two orders of magnitude larger than the magnetic energy differences. These results mostly verified the original idea that drove us to adopt the FM approximation. Most predicted kagome structures remain robust in stability regardless of the magnetic configuration used in the calculations, although Fe-KS competes with T-phase $FeO_2$ depending on their magnetic states.

**Table 2**. List of 12 globally stable kagome monolayers with their classifications and physical properties related to kagome bands near the Fermi level (KB@$E_F$), breathing kagome lattice (BK), $C_{3v}$, bandgap (BG), magnetic moment, and lattice constants. Column Mag. Lists the most stable magnetic configuration among all considered ones that ZZ, dAFM, FM, and LM represent zigzag antiferromagnetic, double stripe antiferromagnetic, ferromagnetic, and local moments (no long-range order), respectively. In the column of magnetic moment, the values before and after the slashes represent the magnetic moments on the metal and oxygen atoms, respectively.

| Group | ML | KB @$E_F$ | BK | $C_{3v}$ | BG (eV) | Mag. | magnetic moment ($\mu_B$) | a (Å) | b (Å) | γ (°) |
|---|---|---|---|---|---|---|---|---|---|---|
| I | Fe-KS | × | √ | × | 0.3 | ZZ | 4.00 / 0.10 | 6.02 | 12.10 | 120.26 |
| | V-KS | × | √ | × | 0.3 | dAFM | 1.90 / 0.00 | 6.07 | 12.29 | 119.57 |
| | Sc-KS | × | √ | √ | × | FM | 0.00 / 0.31 | 6.72 | 6.72 | 120.00 |
| | Y-KS | × | √ | √ | × | FM | 0.00 / 0.34 | 7.32 | 7.32 | 120.00 |
| | Au-KB2 | × | √ | × | 0.9 | LM | 0.78 / 0.16 | 6.59 | 13.15 | 120.06 |
| | Ag-KB2 | × | √ | × | 0.4 | LM | 0.53 / 0.21 | 6.52 | 13.01 | 120.05 |
| | Cu-KB2 | × | √ | × | 0.3 | LM | 0.80 / 0.18 | 5.86 | 11.71 | 120.02 |
| II | Sc-KB1 | √ | √ | × | 2.6 | × | - | 6.90 | 6.90 | 122.86 |
| | Y-KB1 | √ | √ | × | 2.2 | × | - | 7.53 | 7.53 | 123.46 |
| IIIA | $Ta_3O_8$ | √ | × | √ | × | FM | 0.00 / 0.16 | 6.25 | 6.25 | 120.00 |
| | $Nb_3O_8$ | √ | × | √ | × | FM | 0.00 / 0.16 | 6.23 | 6.23 | 120.00 |
| IIIB | $Ir_3O_8$ | √ | × | √ | × | FM | 0.47 / 0.12 | 6.15 | 6.15 | 120.00 |

At the end of the whole workflow, we categorized these 12 monolayers into three groups based on four criteria, that are the presences of kagome bands near the Fermi level (KB@$E_F$), the in-plane $C_{3v}$ rotational symmetry, the bandgap (BG), and the order of magnetic moments, as shown in Table 2. Group-I contains seven kagome monolayers not exhibiting appreciable kagome bands near the Fermi level, which are not our focus in the following discussion. Group-II has two non-magnetic and insulating breathing kagome monolayers, in which Sc-KB1 is a representative. Monolayers in Group-III are metallic standard kagome ferromagnets, which are divided into two subgroups by the distribution of local magnetic moments. Monolayer $Ta_3O_8$ is, on behalf of $Nb_3O_8$, a representative of Group-IIIA where the magnetic moments are contributed by O atoms solely, while $Ir_3O_8$ is the sole member of Group-IIIB where the moments are observable around both Ir and O atoms.

These three representative monolayers, that are Sc-KB1, $Ta_3O_8$, and $Ir_3O_8$, were further discussed in detail. Figure 3(a) plots the band structure of the Sc-KB1 monolayer, in which three kagome bands (in green) below the Fermi level ($E_F$) are labeled as bands B1 to B3. Each Sc atom contributes three valence electrons, preferring a 3+ valency state. Considering the 2- valency of O and the Sc-to-O ratio of 2:3 in Sc-KB1, all valence electrons contributed from Sc are transferred to the unfilled O 2*p* orbitals, resulting in the monolayer being an insulator. The Sc-KB1 monolayer exhibits a pronounced structural distortion, as revealed by the fully relaxed atomic structure shown in Fig. 3(b). In this configuration, the O_up atoms form a breathing kagome lattice, and the Sc atoms adopt a breathing coloring-triangle lattice. This distortion introduces inequality between the two O_up triangles in the breathing kagome lattice and breaks the $C_{3v}$ rotational symmetry within individual O_up triangles, rendering the three O_up atoms within each triangle inequivalent. Both features are characteristic in M-KB1 monolayers. As a result, the inequivalence of the O_up triangles lift the degeneracy of the Dirac cone (in bands B2 and B3) at the K point. Moreover, the broken $C_3$ symmetry in the O_up triangles opens a bandgap between the originally degenerated flat band and Dirac states (bands B1 and B2) at the G point. The wavefunction norm square of band B1 at the G point, mapped onto the relaxed structure in Fig. 3(b), reveals that this flat band (band B1) is primarily derived from O 2*p* orbitals. Bands B1 and B2 are characterized by a $Z_2$ invariant of zero, indicating a trivial topological character, while band B3 intersects with other bands multiple times, making it challenging to unambiguously determine its $Z_2$ invariant.

Figure 3(c) plots the spin-polarized band structure of the $Ta_3O_8$ monolayer, which spin-up bands shown in red and spin-down bands in blue. A set of spin-down kagome bands (highlighted in green) crosses the Fermi level, while their spin-up counterparts (in red) are located approximately 0.5 eV below the Fermi level. In the $Ta_3O_8$ monolayer, three Ta atoms contribute a total of 15 valence electrons, while the eight O atoms can accumulate 16 electrons. This presence of one fewer electron then required to fully occupy the valence orbitals of O atoms leads to metallicity, leaving an unoccupied spin-down band, primarily comprised of O 2*p* orbitals, within the green-colored kagome set. The partially filled O 2*p* orbitals causes local magnetic moments predominately distributed on the O atoms, as illustrated by the mapped spin density in Fig. 3(d). If one oxygen atom in $Ta_3O_8$ is substituted with a halogen atom, effectively doping one additional electron into the monolayer, the flat band (band B1) becomes fully occupied, potential driving a metal to insulator transition. Furthermore, the flat band is characterized by a non-zero Chern number (C=–1), indicating it is topologically nontriviality and suggesting the presence of topologically protected properties, such as the fractional quantum anomalous Hall (FQAH) effect.

The $Ir_3O_8$ monolayer is a comparable but more complicated case than the $Ta_3O_8$ monolayer, as metal *d* orbitals are involved in forming kagome bands near Fermi levels and resulting in the ferromagnetism. As shown in its band structure (Fig. 3(e)), there are two spin-down kagome band sets near the Fermi level, one above (KBS-1, shown in green) and one crossing (KBS-2, in violet). The corresponding spin-up sets reside across and below the Fermi level, respectively. Each spin-down kagome set has a Dirac point at the K point, marked by circles in Fig. 3(e). These Dirac bands are formed from a hybridization of Ir $t_{2g}$-*d* and O 2*p* orbitals, as indicated by the top-views of the wavefunction norm squares of two representative bands from KBS-1 (Fig.3(f)) and KBS-2 (Fig. 3h)). In KBS-1, both the $t_{2g}$-*d* and O 2*p* orbitals are tilted relative to the layer plane, while in KBS-2, the orbitals are oriented perpendicular to the layer plane. The tilted Ir *d* orbitals in KBS-1 interact through 2*p* orbitals of O atoms within an O sublayer (Fig. 3(f) and S12(f)). However, the perpendicularly orientation in *d* and *p* orbitals in KBS-2 facilitates their overlaps through O atoms from both O sublayers (Fig. 3(h), S12(h) and S12(i)).

The magnetism in $Ir_3O_8$ is notably distinct from that in other monolayers. The three Ir atoms in $Ir_3O_8$ contribute 27 valence electrons in total, of which16 are captured by the eight O atoms, leaving 11 electrons to occupy the $t_{2g}$ orbitals of the three Ir atoms. With an on-site Coulomb energy of 3 eV included in the calculations, the $Ir_3O_8$ monolayer exhibits a small spin-splitting and thus prefers a low-spin state. Among these 11 electrons, ten fill in six spin-up and four spin-down states, while the remaining electron partially fills one spin-up and one spin-down state due to their comparable energies. This configuration results in a net magnetic moment of 2 $\mu_B$, as illustrated in Fig. 3(g).

A notable challenge for $M_3O_8$ monolayers is their exceptionally high work functions. Unlike the 4.9 eV work function of Sc-KB1, the work functions of the $Ta_3O_8$ and $Ir_3O_8$ monolayers are 8.1, and 7.5 eV, respectively, which are higher than most of known monolayers. Such high work functions make $M_3O_8$ monolayers strong electron acceptors and less stable. This issue can be mitigated through electron doping, e.g. via element substitution. For instance, substitution of seven O atoms with halogen atoms ($Ir_3OX_7$ (X=F, Cl, Br, I)) dopes seven electrons into the monolayer, transforming it into an insulator with substantially enhanced stability. Further doping with one more electron into the monolayer, e.g. in forms of $Ir_3X_8$ (X=F, Cl, Br, I), fills in one spin-polarized band of the two sets of kagome bands comprised of unoccupied Ir $e_g$ orbitals. This addition electron drives a transition from an insulating to a metallic state, highlighting the tunable electronic properties of these monolayer through controlled substitution.

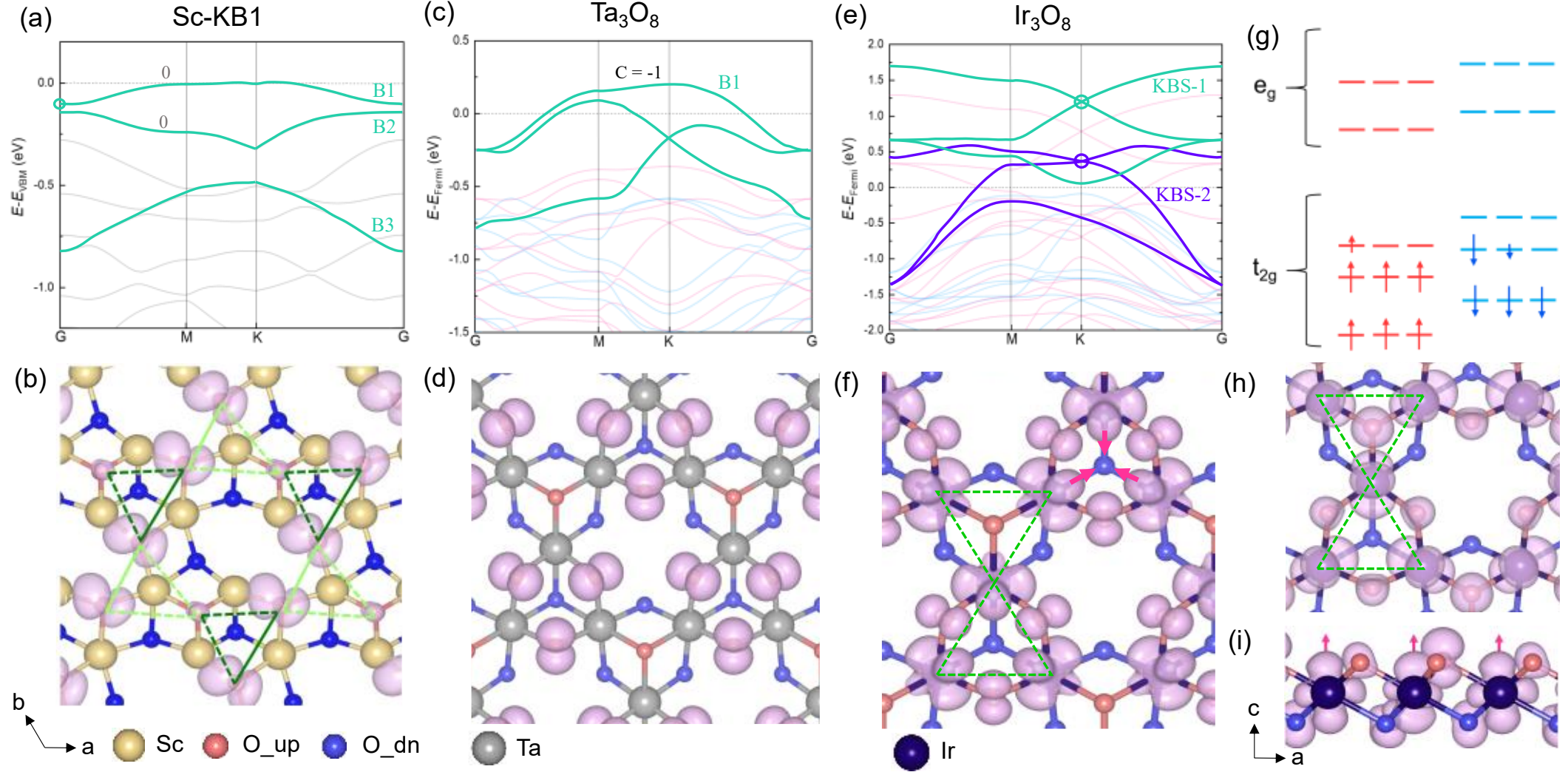

**Fig. 3.** Electronic structure of KB1-Sc, $Ta_3O_8$, and $Ir_3O_8$. (a) Band structure of KB1-Sc. The green lines indicate the kagome bands in KB1-Sc, which are labeled by B1 to B3. The $Z_2$ topological invariants of B1 and B2 are shown as gray numbers. (b) Top view of the wavefunction norm square of band B1 at the G point, indicated by the green hollow circle in (a). The dark and light green triangles represent the two distinct triangles in the breathing kagome lattice. In each green triangles, the two edges of different lengths are denoted by solid and dashed lines, respectively. The isosurface value was set to 0.001 e/Bohr$^3$. (c) Band structure of $Ta_3O_8$. The green lines indicate the kagome bands, with the flat band labeled as B1. (d) Top view of the spin density of $Ta_3O_8$. The isosurface value was set to 0.002 e/Bohr$^3$. (e) Band structure of $Ir_3O_8$. The green and violent lines represent two sets kagome bands in $Ir_3O_8$, labeled as KBS-1 and KBS-2 respectively. (f, h) Top view of the wavefunction norm squares of $Ir_3O_8$ at the K points of KBS-1 (f) and KBS-2 (g). The dashed outlines in (f, h) represent the kagome lattice formed by transition metal atoms. The isosurface value was set to 0.0027 e/Bohr$^3$. (i) Side view corresponding to (h). Pink narrows indicate the orientation of Ir 5*d* orbitals. (g) Schematic diagram of the electron distribution in $Ir_3O_8$. The red (blue) arrows represent spin up (down) electrons, where the large arrows indicate one electron, and the small arrows indicate 0.5 electron.

## 4. Conclusion

In this study, we employed a "1+3" design strategy, combined with high-throughput calculations, to systematically explore thermodynamically stable kagome materials within TMO monolayers. Our investigation identified 12 stable monolayers with diverse magnetic properties, including five ferromagnetic (FM), two antiferromagnetic (AFM), two non-magnetic, and three exhibiting only local magnetic moments. Among these, five monolayers were found to host kagome bands near their Fermi levels, from which Sc-KB1, $Ta_3O_8$, and $Ir_3O_8$ were selected as representatives for detailed analysis. We revealed a clear link between the electronic origins of the kagome bands and the degree of *d* electron localization on the transition metal atoms. When the *d* -electrons are fully transferred to oxygen, the kagome bands originate predominantly from O 2*p* orbitals. Conversely, when the transition metal *d*-orbitals remain partially occupied, the kagome bands emerge as hybrid states involving both metal *d* and oxygen 2*p* orbitals. Our findings demonstrate the versatility and feasibility of the "1+3" strategy for designing kagome lattices and provide valuable insights into the intricate interplay among electronic structure, magnetism, and orbital contributions in $TM_3X_8$ kagome monolayers (TM=transition metals, X=VI-A or VII-A group elements). This study lays the foundation for the theoretical design of kagome monolayers, offering candidate materials for experimental efforts to unlock their distinctive quantum phenomena, such as topological states and correlated electronic properties.

**Data availability statement**

The data that support the findings of this study are openly available in Science Data Bank at https://www.doi.org/XXXXXXX. This statement should be given if some related data have been deposited in Science Data Bank.

**Acknowledgments**

We gratefully acknowledge the financial support from the Ministry of Science and Technology (MOST) of China (Grant No. 2023YFA1406500), the National Natural Science Foundation of China (Grants No. 11974422 and 12104504), the Fundamental Research Funds for the Central Universities, and the Research Funds of Renmin University of China [Grants No. 22XNKJ30 (W.J.) and 24XNKJ17 (C.W.)]. R.L. acknowledges the National Scientific and Technological Innovation Talent Training Program (CAST & MOE) for offering him an opportunity to participate in this research. All calculations for this study were performed at the Physics Lab of High-Performance Computing (PLHPC) and the Public Computing Cloud (PCC) of Renmin University of China.